# Decoupling of Self Diffusion from Viscosity of Supercooled Water: Role of Translational Jump-diffusion

*Shivam Dueby[†], Vikas Dubey[†], Snehasis Daschakraborty[*]*

Department of Chemistry, Indian Institute of Technology Patna, Bihar 801106, India.

AUTHOR INFORMATION

**Corresponding Author**

[*]snehasis@iitp.ac.in

**Abstract:** Some experiments have witnessed increasing decoupling of viscosity from the translational self-diffusion of supercooled water with decreasing temperature. While theory and computer simulation studies indicated the jump translation of the molecules as a probable origin of the above decoupling, a precise quantitative estimation is still lacking. Through a molecular dynamics (MD) simulation study, along with careful consideration of translational jump motion, we have found the most definite proof of increasing relevance of translational jump diffusion in the above decoupling phenomena. By separating out the jump-only diffusion contribution from the overall diffusion of the water, we obtain the residual diffusion coefficient, which remains strongly coupled with the viscosity of the medium at the whole temperature range, including supercooled regime. These new findings can help to elucidate many experimental studies featuring molecular transport properties, where strong diffusion-viscosity decoupling comes into the picture.

There are intriguing properties of supercooled water, including a strong decoupling between its viscosity and the diffusion of the molecules. Some experimental studies[1-3] —including that by Dehaoui et al.[4]— has revealed an increasing decoupling of viscosity $\eta$ from the water translational diffusion coefficient $D_t$ upon cooling. This indicates a gradual breakdown of the Stokes-Einstein (SE) relation ($D_t \propto T/\eta$) with decreasing temperature. In contrast, the rotational diffusion $D_r$ remains coupled with $\eta$ for a wide range of temperature, which implies the validity of the Stokes-Einstein-Debye (SED) relation. Similar decoupling between $D_t$ and $\eta$ was also reported earlier in other molecular glass forming liquids.[5-13] The SE relation is obeyed at sufficiently high temperature, but severely breaks down around $1.3T_g$ ($T_g$ is the glass transition temperature). On the contrary, the rotational diffusion of the molecular glass forming liquid and the medium viscosity remain hydrodynamically coupled even at the temperature very close to $T_g$.

Deeply supercooled liquids have spatially heterogeneous dynamics, which have been confirmed by various experiments (e.g., see Refs. 5,6, 14-17) and computer simulation studies (e.g., see Refs. 18-22). A number of computer simulation studies have indicated that the emerging spatiotemporal heterogeneity in supercooled water and other supercooled liquids has connection with the increasing violation of the SE relation with decreasing temperature.[23-30] Recently, two of us have shown that the rotation assisted translational movement of solvent water around a nonpolar solute induces translational jump-diffusion of a tracer from one solvent cage to another in supercooled water.[23]

Even though the prior studies have implied the pivotal role of translational jump-diffusion for the breakdown of the SE relation in supercooled water, a quantitative estimation of the explicit contribution of the jump-only diffusion $D_{Jump}$ (diffusion due to jump only motion) is still missing. This work is an MD simulation attempt for quantitative estimation of the translational diffusion coefficient due to the translational jump of the molecules and subsequently separates the $D_{Jump}$ from the overall diffusion of the water molecules $D_t$. This allows us to check the coupling of the viscosity with both $D_{Jump}$ and $D_t$ and, thus, obtain more quantitative insight into the role of the translational jump-diffusion of the water molecules for the increasing breakdown of the SE relation with decreasing temperature. The organization of the letter is as follows. We first show the breakdown of the SE relation with decreasing the temperature. Then we perform the jump analysis to obtain the jump-only diffusion contribution to the overall diffusion.

Our simulation box contains 2000 water molecules (modeled by TIP4P/2005 force field[31]). Sec. S1 of the Supplemental Material (SM)[32] details the simulation protocol, the validity of which is evidenced by the excellent agreement of the simulated parameters —density (see Figure S1 of the SM), diffusion coefficient, and viscosity coefficient— with the available experimental values. We calculate $D_t$ for a set of 20 water molecules using the mean square displacement (MSD) route $\lim_{t\to\infty}\langle|r(t)-r(0)|^2\rangle = 6D_t t$. We randomly pick these 20 water molecules from the entire ensemble to keep the movement of this set of water molecules as independent as possible. However, we see that $D_t$ for this set of 20 water molecules is identical with that of the full set of 2000 water molecules. Figure S1 of the SM presents the MSD against time for the whole temperature range. Figure 1a presents the simulated $D_t$ as a function of temperature, which is in excellent agreement with the available experiment[33]. (Both the simulated and the experimental values are listed in Table S1 of the SM.) We found that the simulated data fits quite well with the empirical Vogel-Fulcher-Tamman (VFT)-type relationship, $D_t = D_{t0}\exp[-B/(T-T_0)]$, where $T$ is the temperature, $T_0$ is generally close to the glass transition temperature, and $D_{t0}$ and $B$ are other fitting constants. We obtain the following fitting parameters after regressing the VFT equation onto the simulated $D_t$: $D_{t0} = 3.48 \times 10^{-4}$ cm$^2$/sec, $B = 339.23$ K, and $T_0 = 175.1$ K. These values agree well with the experimental data.[33] We calculate the viscosity $\eta$ of the water using the Green-Kubo relation via stress tensor correlation function. Figure 1b presents an Arrhenius plot of $\eta$ as a function of temperature. Clearly, the simulated $\eta$ matches very well with the measured $\eta$ values at all the temperatures. (Both the simulated and the experimental values [4] are listed in Table S1 of the SM.) The above agreement validates the simulation force field and other parameters. (See the simulation protocol in Sec. S1 of the SM). Regression of the VFT equation for viscosity ($\eta = \eta_0 \exp[B/(T - T_0)]$) onto the simulated data in Figure 1b gives the values of $\eta_0$, B, and $T_0$: $9.82 \times 10^{-4}$ P, 278.64 K, and 180.34 K respectively, which are consistent with the experimental fitting parameters [4].

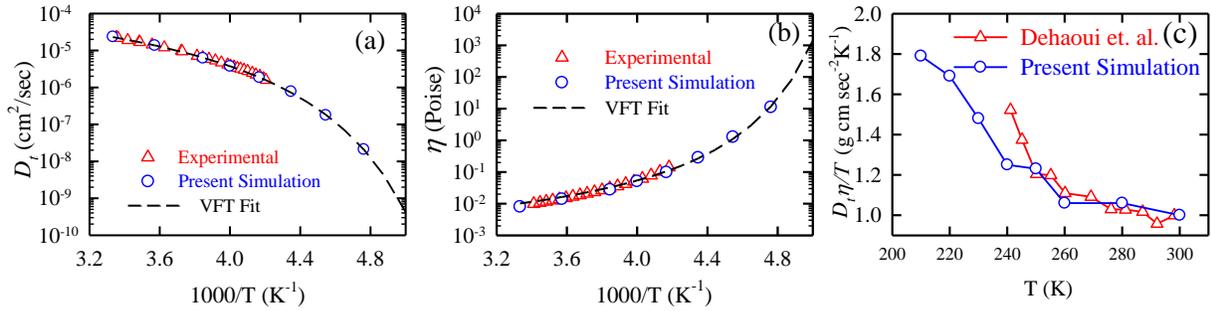

Figure 1. Comparison between the simulated and the measured transport coefficients[33]. Arrhenius plots of simulated and measured translational diffusion coefficient $D_t$ (a), and shear viscosity coefficient $\eta$ (b) as functions of temperature. The simulated data are fitted with the VFT equation in both (a) and (b). (c) Simulated and measured $D_t\eta/T$ as a function of temperature.

We now check the validity of the SE relation using the above simulated $D_t$ and $\eta$ values. The simulated and experimental $D_t\eta/T$ values for the whole temperature range are listed in Table S1 of the SM. Figure 1c exhibits the normalized simulated and experimental $D_t\eta/T$ as functions of temperature, which should be constant if the SE relation holds correctly. Note that the experimental $D_t\eta/T$ values are obtained from the measured $D_t$[31] and $\eta$ values[4]. Normalization of the experimental and the simulated $D_t\eta/T$ values are done with respect to the value at T=300 K. Figure 1c clearly shows the gradual deviation of the normalized simulated $D_t\eta/T$ from unity as the temperature decreases from the room temperature. This indicates an increasing violation of the SE relation —which reaches ~80% — as we decrease the temperature down to 210 K. This is consistent with the available experimental result down to T = 240 K temperature.

Now, we turn our focus on the quantitative jump analysis for estimating the jump-only diffusion coefficient $D_{Jump}$ of the water molecules at all the simulated temperatures. One of the most crucial steps of the analysis is the correct identification of the translational jump occurrence. Two different approaches are available. The first method —based on the displacement of a molecule from its position at the beginning of the trajectory $t$=0— is easier to implement and thus frequently used.[23,34-38] However, this approach has serious problems in the quantitative analysis due to the following reason. This method correctly identifies only those jumps, where the initial and the final positions —the initial position is the position of the molecule just before the jump occurrence, and the final position refers to the new position of the molecule after the jump occurrence— of the jumping molecule are colinear with the position at $t$=0. Inaccuracy increases in the process of identification of the jump occurrence and calculating the jump length with a gradual deviation of the above three coordinates from linearity.[23] The method completely loses its hold when the three coordinates form a right angle triangle. In that case, the jump displacement shows either no peak at all or a very small peak with an intensity too low to detect among the thermal noise. This method, therefore, underestimates the contribution of jump-diffusion to the overall diffusion. A more quantitative method is therefore necessary.

We have used here a more quantitative method, which is similar to that developed by Raptis et al.[39,40] and later used by Araque et al.[41] This method is based on the calculation of the radius of gyration $R_g$ of different segments of the molecular trajectory in three-dimension position coordinate space. The radius of gyration for the particular trajectory segment of length $\Delta t$ (or $n$ number of time steps) is calculated using the following equation.

$$R_g(t, \Delta t) = \sqrt{\frac{1}{n}\sum_{i=1}^{n}[\bm{r}_i(t;\Delta t) - \bm{r}_{CM}(t;\Delta t)]^2}. \quad (1)$$

In eq. 1, $\bm{r}_i(t;\Delta t)$ and $\bm{r}_{CM}(t;\Delta t)$ are, respectively, the position of $i^{th}$ time frame and the center of mass of the trajectory segment of length $\Delta t$. $\bm{r}_{CM}(t;\Delta t)$ is calculated by the following equation.

$$\bm{r}_{CM}(t, \Delta t) = \frac{1}{n}\sum_{i=1}^{n}\bm{r}_i(t;\Delta t) \quad (2)$$

As the diffusion of water increases with the temperature, (see Figure 1a) consideration of the same $\Delta t$ for all the temperatures can lead to unreasonable results. For example, $\Delta t$=10 ps trajectory segment at 210 K temperature spreads in space much less compared to that at 300 K. In order to avoid this aparant incosistancy, we choose different $\Delta t$ for different temperatures. It is observed that consideration of $\Delta t$ as the characteristic time $t^*$ —when the non-Gaussian parameter $\alpha_2(t)$ is maximal— works properly at all the temperatures.[41] Also, the peak of $\alpha_2(t)$ corresponds the highest heterogeneous dynamics of the molecules at time $t^*$.[42,43] We calculate $\alpha_2(t)$ using the following equation:

$$\alpha_2(t) = 3\langle r^4(t)\rangle/5\langle r^2(t)\rangle^2 - 1 \qquad (3).$$

Figure 2a presents $\alpha_2(t)$ against time for the whole temperature range (T=210 K to 300 K). The maxima of $\alpha_2(t)$ increases with decreasing temperature. This indicates the increase of dynamical heterogeneity due to the decrease of temperature. Also, $t^*$ increases from ~1 ps to ~550 ps (see Table S2 of the SM) while the temperature decreases from T=300 K to 210 K. Therefore, the length $\Delta t$ of the trajectory segment —the input in equations 1 and 2— increases with decreasing temperature.

Once we divide the trajectory into multiple segments of length $\Delta t$ each, we calculate $R_g$ for all the trajectory segments separately. The distance traversed by the molecule in a trajectory segment $\lambda(t)$ can be calculated from the formula, $\lambda(t) = 2R_g$.[39-41] We note that the translational jump occurrences are not ubiquitous in all these trajectory segments. At this point, we need an efficient method in order to correctly identify the translational jump segments. For this, we use the same method adopted by Araque et al. The method uses the self-part of the van Hove correlation function $G_S^{simu}(r,t)$, which is calculated from the following equation[42,43]:

$$G_S^{simu}(r,t) = \frac{1}{N}\langle \sum_i^N \delta(r - |\boldsymbol{r}_i(t_0) - \boldsymbol{r}_i(t)|)\rangle_{t_0} \qquad (4).$$

$G_S^{simu}(r,t)$ deviates from the Gaussianity ($G_S^{theo}(r,t) = [3/2\pi\langle r^2(t)\rangle]^{3/2}\exp[-3r^2/2\langle r^2(t)\rangle]$) [40,41] at the most at time $t^*$. Both the $G_S^{simu}(r,t)$ and $G_S^{theo}(r,t)$ are plotted in Figure 2b for T=210 K and in Figure S2 of the SM for the rest of the temperatures. $G_S^{simu}(r,t)$ crosses $G_S^{theo}(r,t)$ at two characteristic $r$ values; $r_1$ and $r_2$, by which we define the jump and the cage trajectories respectively. At the smaller $r$ limit ($r < r_1$) the actual displacements of the water molecules is less than the theoretical value obtained from $G_S^{theo}(r,t)$. Therefore, the trajectory segment of length $\lambda(t) < r_1$ is a cage trajectory. On the other hand, at the larger $r$ limit ($r > r_2$) the actual displacement of the water molecule is larger than the displacement obtained theoretically from $G_S^{theo}(r,t)$. (Table S2 of the SM presents the numerical values of $r_1$ and $r_2$ for the different temperatures.) Therefore, the trajectory segments, where traversed distance by the molecule $\lambda(t)$ > $r_2$, is categorized as jump trajectory.[41] Figure 2c exhibits $\lambda(t)$ as a function of time for one water molecule at T=210 K as the representative temperature. A peak, whose intensity crosses the distance $r_2$, represent a translational jump trajectory. Conversely, we define a trajectory segments as cage trajectory, which has $\lambda$ value less than the cutoff $r_1$. Here, the cage trajectory refers to the rattling motion of a molecule inside the solvent cage plus the effective translation of the overall solvent cage. Figures S3 and S4 of the SM present several examples of cage and jump segments for two representative temperatures. It is evident from the representative examples that the above protocol for identifying the jump and cage trajectory is working exactly as we expected that the identified jump trajectory segments consist of sudden change of position of the molecule in between two cage rattling and that the wrapping-up of the trajectory is observed in cage trajectory segments.

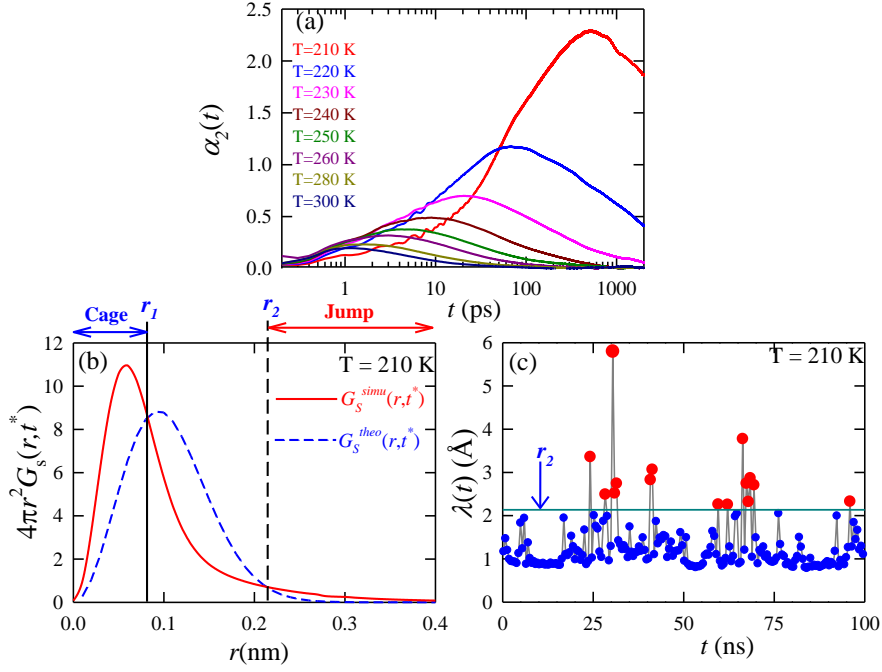

Figure 2. (a) Non-Gaussian parameter $\alpha_2(t)$ as a function of time for all the temperatures studied. The time $t^*$ for the maximum value of $\alpha_2(t)$ are listed in Table S2 of the SM. (b) The self-part of the van Hove correlation functions (solid line) and the corresponding ideal Gaussian distribution ($G_S^{theo}(r,t)$) (dashed line) at time $t^*$ when $\alpha_2(t)$ is maximum at T=210 K. The similar plots for other temperatures are presented in Figure S2 of the SM. (c) The distance traversed by one water molecule, $\lambda(t)$, in a individual trajectory segment, which is centered at time $t$. The horizontal line indicates the cutoff distance $r_2$. The red circles, which are above the cutoff distance $r_2$, represent the jump trajectory segments.

As we have correctly identified the jump and the cage trajectory segments, we now calculate the jump-only translational diffusion coefficient $D_{Jump}$ using the following equation:

$$D_{Jump} = \frac{1}{6}\nu\lambda_J^2 \qquad (5).$$

Here, $\nu$ is the frequency of the translational jump occurrence of the water molecules (number of jump occurrences $n_{Jump}$ /number of water molecules /the full production trajectory length 100 ns) and $\lambda_{Jump}$ is the average jump length, which is obtained from $R_g$ using the formula, $\lambda_{Jump} = 2R_g$ for the jump trajectory segments only. We have listed, in Table S3 of the SM, the numerical values of $n_{Jump}$, $\nu_{Jump}$ and $\lambda_{Jump}$ for the whole temperature range. Using the above numerical values, we calculate $D_{Jump}$ of the water molecules at all the temperatures, which are listed in Table S4 of the SM. Figure 3a exhibits the percentage contribution of the $D_{Jump}$ to the overall diffusion $D_t$ of the water molecules as a function of temperature. The contribution increases with decreasing temperature and reaches to more than 50% of $D_t$ at T=210 K. However, this profound increase does not necessarily mean a similar increase of the jump frequency. Table S3 of the SM clearly shows that the jump trajectory segments are only c.a. 0.9% of the total trajectory segments at T=300 K and increases up to only c.a. 4.4 % at T=210 K. Therefore, approximately 1 jump in 20 trajectory segments contribute more than 50% of the overall diffusion of the water molecules at T=210 K. In other words, a small fraction of the jump trajectory contributes to a large fraction of the overall diffusion at T=210 K.

The above method has also been used to obtain the cage diffusion coefficient $D_{Cage}$, which parameterizes the overall translational movement of the water cluster consisting of the tagged and cage water molecules. We now check the validity of the SE relation using the calculated $D_{Jump}$, $D_{Cage}$, and the simulated $\eta$ of the medium. Figure 3b displays the normalized $D_{Jump}\eta/T$ and $D_{Cage}\eta/T$ values as functions of time. While, $D_{Cage}\eta/T$ remains almost insensitive to the temperature, $D_{Jump}\eta/T$ increases rapidly with decreasing the temperature. The value of the latter increases by ~22 times as the temperature is decreased from T=300 K to 210K. Therefore, as expected, $D_{Jump}$ is completely decoupled from the viscosity almost at all the temperatures.

Lastly, using the jump-only and the overall diffusion coefficients one can obtain the residual diffusion coefficients $D_{Res}$ using the equation:

$$D_{Res} = D_t - D_{Jump} \tag{6}$$

We have calculated $D_{Res}$ for all the temperatures and listed their numerical values in Table S4 of the SM. We now check the validity of the SE relation by checking the effect of the temperature on the numerical value of $D_{Res}\eta/T$. Figure 3c plots $D_{Res}\eta/T$ as a function of temperature. Very interestingly, unlike the $D_t\eta/T$, which continuously increases with decreasing the temperature, the $D_{Res}\eta/T$ value almost remains constant.[1-4] This indicates that once the translational jump-only diffusion is separated out from the overall diffusion coefficient, the residual diffusion perfectly couples with the viscosity and therefore follows the SE relation. Therefore, the jump-diffusion of the molecules is the central origin for the observed decoupling of the molecular diffusion from the viscosity of the medium. This is one of the key results of this work as it categorically proves the hypothetical concept that the origin of the well-known diffusion-viscosity decoupling in supercooled water (liquid) is the translational diffusion, larger than expected. Two of us have previously shown the mechanism of these translational jumps in great details and the crucial role of the synchronization between the translational and the rotational motion of the solvent water molecules for inducing these jump events.

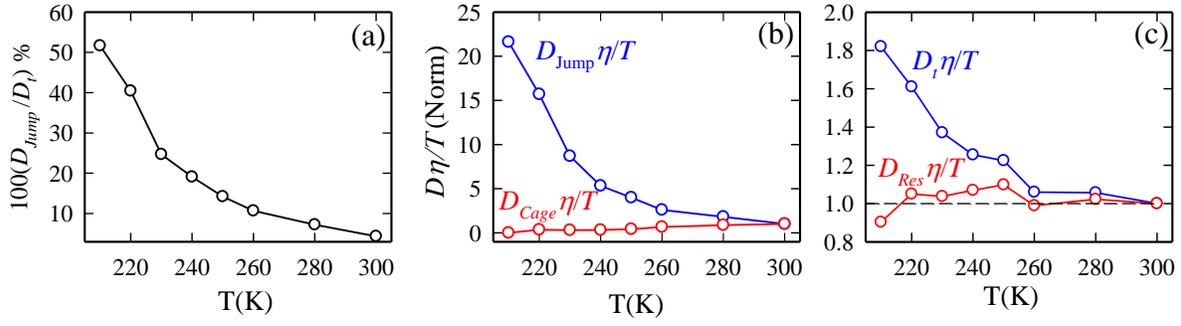

Figure 3. (a) Contribution of the jump-only diffusion coefficient $D_{Jump}$ (in percentage) to the overall diffusion coefficient $D_t$ of the water molecules as a function of temperature. Temperature-dependent coupling of viscosity with (b) translational jump diffusion coefficient $D_{Jump}$, the cage diffusion coefficient $D_{Cage}$, (c) the overall translational diffusion coefficient $D_t$, and the residual diffusion coefficient $D_{Res}$ ($D_t$-$D_{Jump}$) at the temperatures studied.

In conclusion, we have presented an MD simulation analysis, detailing the quantitative role of translational jump-diffusion on the increasing decoupling of viscosity from the translational diffusion with decreasing the temperature. By careful consideration of the translational jump trajectories of the water molecules, we have calculated the jump-only diffusion coefficient $D_{Jump}$. As the temperature decreases, the contribution of $D_{Jump}$ to the overall diffusion increases. Once we separate out the $D_{Jump}$ from the overall diffusion coefficient of water $D_t$, we obtain the residual diffusion coefficient $D_{Res}$. While $D_{Jump}$ intensely decouples from the viscosity $\eta$, $D_{Res}$ stays coupled strongly with $\eta$ at all the temperatures. This is an absolutely clear evidence of the translational jump-diffusion of the molecules as the key origin for the observed decoupling of viscosity from the translational diffusion. These new findings can help in elucidating many experimental studies featuring molecular transport properties in more complex chemical and biological environment, where strong diffusion-viscosity decoupling comes into the picture. In addition, a modified version of the above methodology can be used for calculating the rotational jump-only diffusion coefficient for non-associated liquid. This would generalize the existing rotational jump model for liquid water.[44,45]

## AUTHOR INFORMATION

**Corresponding Author**

[*]snehasis@iitp.ac.in

**Author Contributions**

[†]Shivam Dueby and Vikas Dubey contributed equally.

**Notes**

The authors declare no competing financial interest

## ACKNOWLEDGMENT

We acknowledge Prof. James T. Hynes, Prof. Jannis Samios, Prof. Claudio E. Margulis, and Dr. Sandipa Indra for fruitful discussion and IIT Patna for the computational facility. Shivam and Vikas acknowledge IIT Patna for their research fellowships. We are thankful to Archita for assistance in graphical designing.